\documentclass[aps,twocolumn,showlabels,showrefs,amsmath,amssymb,pre,superscriptaddress,floatfix,colors]{revtex4-1}

\usepackage{graphicx}
\usepackage{dcolumn}
\usepackage{bm}
\usepackage{amssymb}
\usepackage{multirow}
\usepackage{color}
\usepackage{ulem}
\usepackage{float}
\usepackage{appendix}
\begin{document}

\title{Unzipping of knotted DNA via nanopore translocation}

\date{\today}


\author{Antonio Suma}
\affiliation{Dipartimento  di  Fisica,  Universit\`a  degli  Studi  di  Bari,  via  Amendola  173,  Bari,  I-70126,  Italy}
\affiliation{Institute for Computational Molecular Science, Temple University, Philadelphia, PA 19122, USA }
\author{Cristian Micheletti}
\affiliation{Scuola Internazionale Superiore di Studi Avanzati (SISSA), Via Bonomea 265, 34136 Trieste, Italy}

\begin{abstract}
DNA unzipping by nanopore translocation has implications in diverse contexts, from polymer physics to single-molecule manipulation to DNA-enzyme interactions in biological systems. Here we use molecular dynamics simulations and a coarse-grained model of DNA to address the nanopore unzipping of DNA filaments that are knotted. This previously unaddressed problem is motivated by the fact that DNA knots inevitably occur in isolated equilibrated filaments and {\it in vivo}. We study how different types of tight knots in the DNA segment just outside the pore impact unzipping at different driving forces. We establish three main results. First, knots do not significantly affect the unzipping process at low forces. However, knotted DNAs unzip more slowly and heterogeneously than unknotted ones at high forces. Finally, we observe that the microscopic origin of the hindrance typically involves two concurrent causes: the topological friction of the DNA chain sliding along its knotted contour and the additional friction originating from the entanglement with the newly unzipped DNA. The results reveal a previously unsuspected complexity of the interplay of DNA topology and unzipping, which should be relevant for interpreting nanopore-based single-molecule unzipping experiments and improving the modeling of DNA transactions in vivo.
\end{abstract}

\maketitle

\noindent

\section*{Introduction}
A series of advancements in pore translocation setups have brought this single-molecule technique to the forefront of numerous applications, far exceeding the originally envisioned purpose of sequencing nucleic acids~\cite{kasianowicz1996characterization,palyulin2014polymer,deamer2016three}. Recent applications include advanced molecular sensing~\cite{rahman2019demand,wang2021slowing,leitao2023spatially}, out-of-equilibrium stochastic processes~\cite{kantor2004anomalous,grosberg2006long,sarabadani2018theory,suma2023nonequilibrium}, RNA unfolding~\cite{bandarkar2020nanopore,suma2020directional}, protein sequencing~\cite{asandei2020nanopore,yu2023unidirectional}, and probing of intra- and inter-molecular entanglement~\cite{huang2008translocation,rosa2012topological,suma2015pore,narsimhan2016translocation,plesa2016direct,suma2017pore,marenda2017sorting,caraglio2017driven,weiss2019hydrodynamics,caraglio2020translocation,rheaume2023nanopore}.

One of the most exciting avenues for nanopore translocation is probing the structure and function of biological polymers. A notable example is offered by exonuclease-resistant RNAs (xrRNAs)~\cite{pijlman2008highly,chapman2014structural,akiyama2016zika,macfadden2018mechanism,slonchak2020zika,vicens2021shared}. These modular elements, consisting of only a few dozen nucleotides, are located at the 5$^\prime$ end of the RNA genome of flaviviruses and are responsible for infections such as  Zika, dengue, and yellow fever~\cite{slonchak2018subgenomic}. xrRNAs are distinguished by their unique and diverse functional responses when pulled through the lumen of enzymes that process nucleic acids. Specifically, xrRNAs resist degradation by exonucleases that translocate nucleic acids from the 5$^\prime$ end. However, they can be processed by replicases and reverse transcriptases, which translocate RNAs from the 3$^\prime$ end.

A mechanistic explanation for this behavior was provided by the theoretical and computational study of ref.~\cite{suma2020directional}, where a pore translocation setup, mimicking the action of processive enzymes, was used to unzip xrRNAs from both ends. The study, further supported by later work~\cite{becchi2021rna,niu2020},
reported that the short and yet heavily entangled structure of xrRNAs, which includes several pseudoknots~\cite{akiyama2016zika}, contributes to a strongly directional translocation response. Pulling xrRNAs from the 5$^\prime$ end causes the molecule to close in on itself and resist further unzipping, explaining its resistance to exonucleases; conversely, when translocated from the 3$^\prime$ end, the molecule progressively unzips and falls apart, explaining its susceptibility to replicases and helicases/reverse transcriptase~\cite{suma2020directional}.

Differently from RNAs, double-stranded DNA (dsDNA) filaments are usually well described by general polymer models with torsional and bending rigidity~\cite{chirico1994kinetics,klenin1998brownian,vologodskii1994supercoiling}. Although dsDNA does not form the complex architectures typical of RNAs, it can become knotted due to its spontaneous dynamics, both in bulk and under confinement~\cite{rybenkov1993probability,arsuaga2002knotting,marenduzzo2009dna}. Additionally, dsDNA filaments can become knotted through the actions of type II topoisomerases, which perform strand crossings that can potentially alter the topological state of DNA, establishing a homeostatic level knotting that needs to be tightly regulated to avoid detrimental consequences for living cells~\cite{portugal1996t7,rybenkov1997simplification,olavarrieta2002dna,deibler2007hin,valdes2018dna,valdes2019transcriptional}.

The emergence of DNA knots, be they formed spontaneously or introduced by topoisomerases, has been traditionally based on gel electrophoresis~\cite{droge19926,trigueros2001novel,valdes2019quantitative}. Such setups harness the different hindrances experienced by molecules with different knot types when moving through the gel mesh. Its main limitation regards the maximum length to which it can be practically applied, which is of the order of 10kb.

\begin{figure}
\centering 
\includegraphics[width=1\columnwidth]{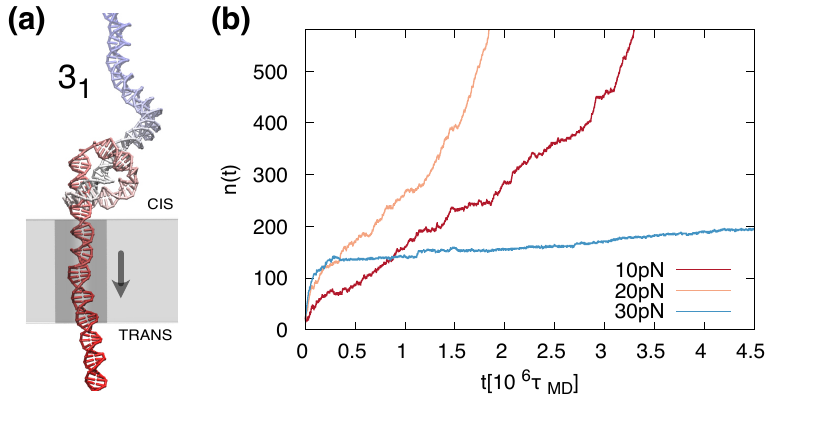}
\caption{(a) Snapshot of a trefoil ($3_1$) knotted dsDNA translocating through a 4.25nm-wide pore, allowing for the passage of a single double strand, thus blocking the knot. The  total translocating force is 30pN, sufficient to maintain the knot in a tight state near the pore entrance. (b) Number of translocated base pairs as a function of time for a $3_1$-knotted dsDNA chain at three different translocating forces. The translocation velocity increases from 10pN to 20pN and then drops dramatically at 30pN, due to the topological friction associated to the strong tightening of the knot. }
\label{Fig1}
\end{figure}

Recent breakthroughs have opened the possibility of overcoming this practical limit by resorting to pore translocation setups~\cite{plesa2016direct,suma2017pore,sharma2019complex}. Suitable choices of the pore diameter allow for translocating the DNA knots and reveal their passage from the drop of the ionic current, which depends on the obstruction of the pore caused by the passing knotted region and involves at least three dsDNA strands. While the technique may not be sensitive to the knot type and knot size~\cite{suma2017pore}, it allows for probing the so-called topological friction~\cite{rosa2012topological,suma2015pore}. The latter can be revealed by using pores sufficiently narrow that only one dsDNA filament can pass through, causing the knot to remain localized at the pore entrance, hindering the translocation of the remainder of the filament that has to slide along the contour of the knotted region to pass through.
In such a setup, the hindrance to translocation can depend on the knot type and the driving force~\cite{rosa2012topological,suma2015pore,narsimhan2016translocation}. Increasing the driving force makes the knots tighter, enhancing the friction to the point that the translocation process can even be stalled indefinitely, Fig.~\ref{Fig1}.

At the same time, dsDNA typically undergoes another type of {\it in vivo} transaction operated by, e.g., ~helicases, namely unzipping. In the pore translocation setup, this effect can be mimicked by reducing the pore diameter so that only one strand of the DNA duplex can pass and is harnessed for fast and reliable genome sequencing~\cite{gundlach2012,Loose2018}.
This interesting out-of-equilibrium setup has been used before to explore fundamental aspects of the equilibrium thermodynamics~\cite{dudko2008theory}, from the sequence-dependent free energy profile~\cite{huguet2010single} of unzipping to base pairing~\cite{suma2023nonequilibrium} to the dynamical regimes appearing at different forces~\cite{suma2023nonequilibrium}, which differ considerably to those occurring without unzipping both in terms of typical translocation times and scaling behavior~\cite{palyulin2014polymer,keyser2021,suma2023nonequilibrium}.

The examples above underscore three key points.
First, the structural features of nucleic acids include physical entanglements, which can have complex and significant functional reverberations {\it in vivo}.
Second, pore translocation setups are indispensable tools for mimicking the action of enzymes and probing the structural response of nucleic acid tangles at the single-molecule level.
Third, the external control afforded by translocation setups, such as varying pore size and force application protocols (constant, time-ramped, oscillating), provides an ideal context for understanding the microscopic basis of the observed unzipping responses. This understanding offers vital clues for decoding how nucleic acid architecture informs translocation.

One open problem that intersects all three aspects above is understanding how the statistically inevitable presence of knots can interfere with DNA unzipping by translocation. Studies have yet to be conducted on this process, which is qualitatively different from translocating knotted DNA without unzipping. For this reason, the insights gleaned from the pore translocation of knotted DNA cannot be directly applied to the unzipping scenario. This leaves fundamental questions about the unzipping of knotted dsDNA unanswered, such as: (i) how large must the driving force be to keep the knot tight at the pore entrance and prevent it from diffusing along the chain, (ii) what is the force-dependent topological friction, and (iii) how does this friction depend on the type of knot?

Here, we address these questions with molecular dynamics simulations of a coarse-grained DNA model, oxDNA2~\cite{oxdna2,oxdna3}. We first consider the reference case of the nanopore unzipping of unknotted DNAs, and study their translocation compliance at different forces. Next, we turn to knotted DNAs and discuss how the unzipping speed varies with knot type and applied force. Finally, we address the complementary aspect, namely how unzipping by translocation affects the knotted region, particularly its length and contour dynamics.

Notably, we did not observe significant effects related to knots at pulling forces of 50pN, which is of the same order as the forces that can be generated by molecular motors~\cite{smith2001bacteriophage}. The results are suggestive that topological entanglement may not significantly interfere with {\it in vivo} DNA unzipping operated by enzymes.
However, the interplay of topology and unzipping is significantly different at 100pN and larger forces, with major effects on the translocation process and knot sliding dynamics.

\section*{Results}

To study the nanopore unzipping of knotted DNA filaments, we applied Langevin molecular dynamics simulations to 500bp-long DNA filaments described with the oxDNA2 model~\cite{oxdna1,oxdna2,oxdna3}, a coarse-grained DNA representation with interactions parameters tuned to reproduce phenomenological data for DNA properties and interactions, including base pairing, stacking, and twist-bend couplings.

\begin{figure*}
\centering 
\includegraphics[width=2\columnwidth]{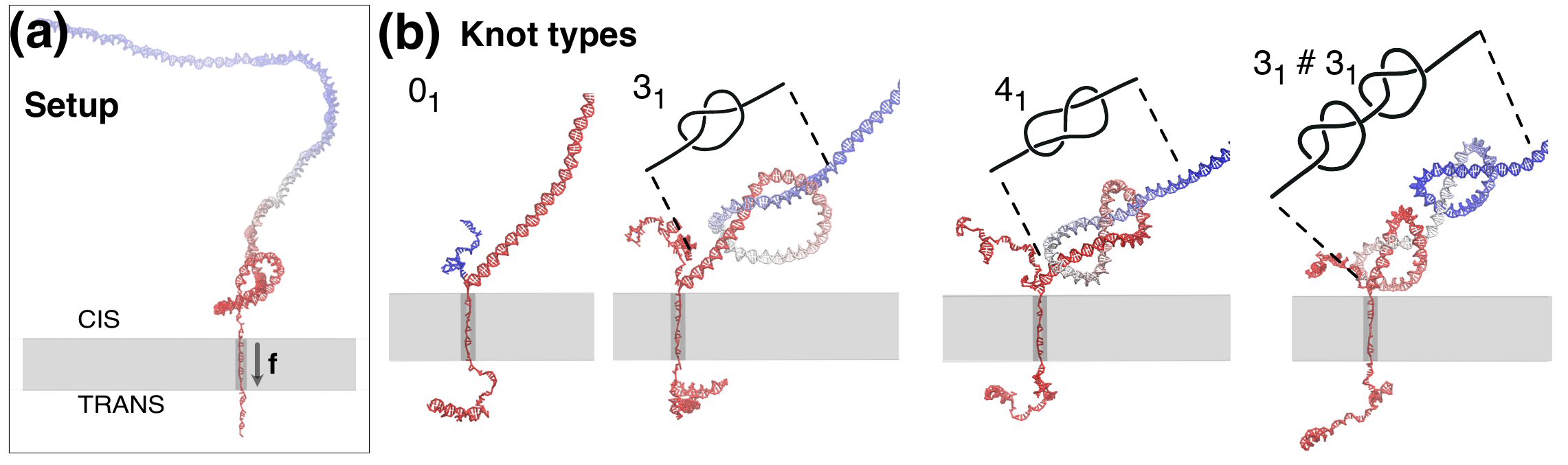}
\caption{(a) Schematic illustration of the initial setup: an unknotted, equilibrated filament is attached to a lead consisting of a tightly-knotted double-stranded segment plus a single-stranded one pre-inserted into a cylindrical pore embedded in a slab.
(b) Configurations of 500bp-long DNA filaments during the simulated translocation-driven unzipping. The four snapshots are close-ups of the system near the pore and illustrate the different considered topologies: unknot ($0_1$), trefoil ($3_1$), figure-of-eight ($4_1$) and the composite granny knot ( $3_1\#3_1$). }
\label{Fig2}
\end{figure*}

The initial states were prepared from five different equilibrated (Monte Carlo generated) conformations of the 500bp filaments. The five conformations were all unknotted because the 500bp contour length, corresponding to about ten DNA persistence lengths, is too short for significant spontaneous knotting in equilibrium\cite{rybenkov1993probability,tubiana2013spontaneous,uehara2019bimodality}.
The 500bp-long filaments were next attached to leads that consisted of a double-stranded knotted region with $3_1$, $4_1$, and $3_1\#3_1$ topology  -- the knotted region was omitted for unknotted ($0_1$) case --  plus a 40-base long single-stranded stretch, pre-inserted into a pore; see Fig.~\ref{Fig2}. The translocation process was driven by pulling the nucleotides inside the pore with a total longitudinal force, $f$ of 50pN, 100pN and 150pN. The pore diameter, 1.87nm, was chosen small enough that only a single DNA strand can pass through it, causing translocating DNAs to unzip.

\subsection{Nanopore unzipping of unknotted DNA}

Fig.~\ref{Fig3}a illustrates, for reference, the translocation response of unknotted DNA filaments. The traces show the number of translocated nucleotides as a function of time, $n(t)$, for five independent trajectories at each indicated force. Note that traces start at about 40, corresponding to the length of the single-stranded DNA (ssDNA) segment of the lead that is already threaded inside the pore at $t=0$.

The traces at $f=50$pN have an overall linear appearance, indicative of an approximately constant unzipping velocity. However, the traces at the two largest forces, 100 and 150pN, deviate noticeably from linearity. The convexity, or upward curvature of the late part of traces ($n(t) >300$), indicates that the average translocation speed increases in the second half of the translocation.

The translocation/unzipping speeds vary significantly across the forces. For comparison, average translocation times were computed at the 400 translocated bases mark, a convenient reference given the graphs' range in Fig.~\ref{Fig3}.  The  average times are equal to $3.0\cdot 10^6, 6.9\cdot 10^5$ and $3.2\cdot 10^5 \tau_{MD}$ for $f=50$, 100, and 150pN, respectively. In particular, we note that the above translocation/unzipping times do not follow the inverse force relationship expected for simple dissipative processes. Specifically, a twofold force variation from 50 to 100pN results in an order of magnitude change in unzipping time.

The results parallel and expand those reported in ref.~\cite{suma2023nonequilibrium}, where data for the out-of-equilibrium unzipping process of dsDNA were used within a framework for stochastic processes to reconstruct the free-energy profile of single base-pair formation. In that context, it was found that the unzipping process proceeded at relatively constant velocity for forces below $\sim60$pN and could be modeled as a drift-diffusive process. At the same time, progressive speed-ups during translocation were observed at larger forces associated with an anomalous dynamics regime. By modelling the unzipping as a stochastic process in a one-dimensional tilted washboard (periodic) potential, it was shown that 60pN force corresponded to lowering the barrier to unzip a base-pair to a value where advective transport becomes relevant over diffusion~\cite{suma2023nonequilibrium}. Additionally, we recall that DNA undergoes significant structural deformations at about this same force when mechanically stretched~\cite{smith1996overstretching}.  Thus, the crossover from linear to non-linear translocation/unzipping observed upon increasing $f$ from 50pN to 100pN is consistent with other qualitative changes of DNA properties in the same force range.

\subsection{Nanopore unzipping of $3_1$-knotted DNA}

\begin{figure}
\centering 
\includegraphics[width=1\columnwidth]{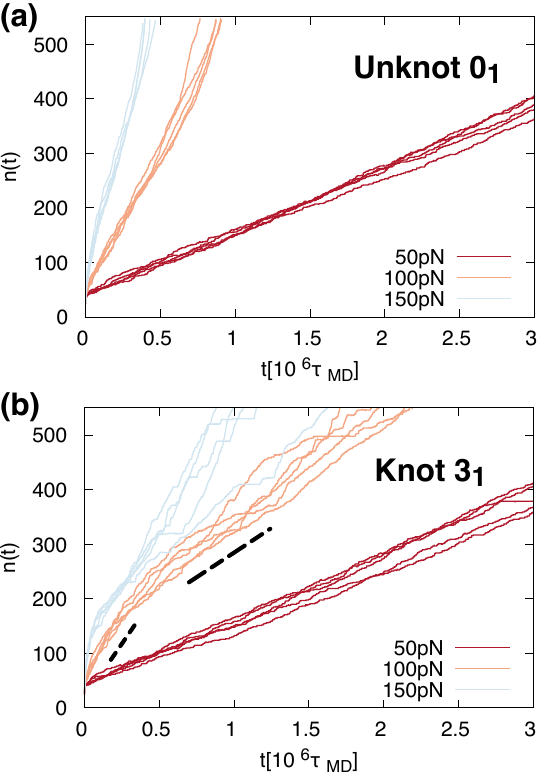}
\caption{Number of translocated nucleotides, $n$, as a function of time, $t$ for dsDNA filaments that (a) are unknotted and (b) have a $3_1$ knot; see Fig.~\ref{Fig2} and methods. The traces are for pulling forces of 50, 100, and 150pN, with five independent trajectories for each case. The dashed lines highlight two distinct velocity regimes in the 100pN trajectories, a feature also present in some of the 150pN traces.}
\label{Fig3}
\end{figure}

The force-dependent translocation response is dramatically changed when the unknotted lead is replaced by a knotted one, even when the topology is the simplest non-trivial one. This emerges by inspecting Fig.~\ref{Fig3}b, which shows the unzipping traces for DNA strands starting with a moderately tight trefoil-knotted ($3_1$) lead.

The comparison of the two panels in Fig.~\ref{Fig3} clarifies that at $f=50$pN, the unzipping of knotted and unknotted chains proceed almost undistinguishably. The average unzipping velocities of the two sets of traces are compatible within statistical uncertainty, $1.309\pm 0.028\cdot 10^{-4} \tau^{-1}_{MD}$ for $0_1$ and $1.288\pm 0.036\cdot 10^{-4} \tau^{-1}_{MD}$  for $3_1$. The main perceived difference is the spread of the five traces, which is larger for the knotted cases.

However, increasing the force to 100pN or more causes the unzipping of knotted chains to proceed more slowly and heterogeneously than unknotted DNAs. For $f=100$pN, the relative slowing down of the average velocity is approximately twofold, and the same holds for the largest considered force, $f=150$pN.

In addition, two different regimes are discernible, highlighted by the dashed lines for the $f=100$ case. Initially, the trefoil-knotted filament unzips at the same rate as the knotted ones. Beyond this regime, which applies to the first 200bp, the process slows down noticeably while also becoming more heterogeneous. An analogous effect is found for the $f=150$pN case, but with the important difference that the transient where the velocity is the same as in the unknotted case has a shorter duration and covers fewer base pairs (150). As we discuss later, the change in velocity is a consequence of the force-induced tightening of the knot near the pore entrance, which adds a significant hindrance - also termed topological friction - to the translocation process.

\subsection{Effect of knot topology on DNA unzipping}

We additionally considered leads with figure-of-eight ($4_1$) and granny ($3_1\#3_1$) knots to extend the range of topological complexity beyond the trivial ($0_1$) and trefoil ($3_1$) knot types. As a conventional measure of knot complexity we consider the crossing number, corresponding to the minimum number of crossings in the simplest possible non-degenerate projection. This complexity measure equals 0, 3, 4, and 6 for the $0_1$, $3_1$, $4_1$, and $3_1\#3_1$ knots, respectively.

\begin{figure*}
\centering 
\includegraphics[width=2\columnwidth]{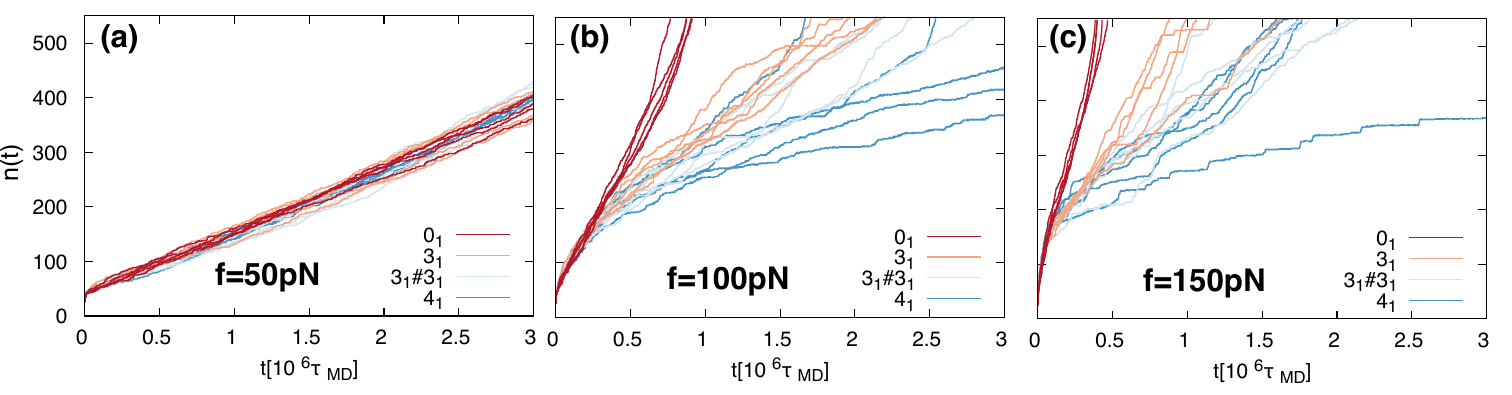}
\caption{Number of translocated nucleotides, $n$, as a function of time, $t$, for DNA filaments with different knot types and at different driving forces, as indicated. The traces of five independent trajectories are shown for each case.}
\label{Fig4}
\end{figure*}

The unzipping traces for all topologies are shown in Fig.~\ref{Fig4}. We stress that we purposely attached the same set of equilibrated 500bp-long dsDNA conformations to the battery of differently knotted leads. With this choice, emerging systematic differences across the different topologies can be directly ascribed to the different knotted states of the lead and not to other effects, such as the initial DNA conformation on the {\it cis} side.

\begin{figure*}
\centering 
\includegraphics[width=2\columnwidth]{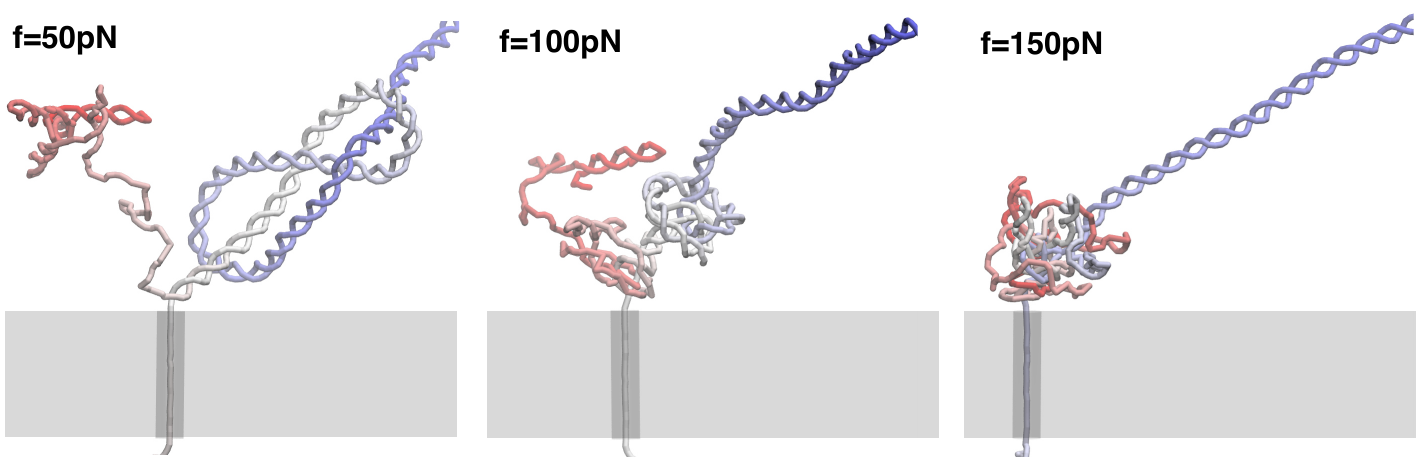}
\caption{Typical conformations of a $4_1$-knotted dsDNA filament at intermediate stages of translocation and increasing driving force, 50, 100, 150pN. At the two largest forces one observes knot tightening and the wrapping of the {\it cis} unzipped strand around the dsDNA region proximal to the pore.}
\label{Fig5}
\end{figure*}

The data in panel (a) show that all traces are well superposed and consistent with an approximate linear (constant velocity) behavior at the lowest considered force, $f=50$pN. This result confirms the earlier observation that the unzipping response is mainly independent of the knotted state at sufficiently small $f$ (Fig.~\ref{Fig3}).

The data in panels (b) and (c), which refer to $f=100$ and $150$pN, respectively, are consistent with those of the trefoil knot case (Fig.~\ref{Fig3}), too, in that the unzipping proceeds practically identically for all topologies of an initial segment spanning 200bp at $f=100$pN and 100bp at $f=150$pN. Beyond this point, the unzipping slows down for all non-trivial knot types.
At $f=100$pN, we observe that the highest unzipping hindrance is offered by the $4_1$ knot, followed by the composite $3_1\#3_1$ knot, and the $3_1$ and $0_1$ topologies. We recall that $3_1\#3_1$  knot has the highest nominal complexity in the considered set, and yet it is not associated with the slowest unzipping at $f=100$pN, which is noteworthy. However, at 150pN, the $3_1\#3_1$ and $4_1$ knots offer comparable hindrance, while the unzipping of the $3_1$ case is faster and that of the unknot $0_1$ remains the fastest.

The findings can be interpreted in terms of previously published results on the translocation - without unzipping - of knotted chains of beads~\cite{suma2015pore}. For such a system, it was shown that each prime knotted component behaves as a dissipative structural element that interferes with the mechanical tension propagating to the chain remainder by significantly reducing it. Without unzipping, the translocation velocity for the case of concatenated trefoil knots ($3_1\# 3_1$) was mainly defined by the force dissipation within the first $3_1$-knotted component, which is less complex than the $4_1$ knot. This observation helps rationalize that in specific force regimes, the hindrance of the $3_1\# 3_1$ case can be intermediate to the $3_1$ and $4_1$ ones.

The results of Figs.~\ref{Fig3} and \ref{Fig4} establish two points. First, the effects of DNA knots on the unzipping process are negligible, up to forces of at least 50pN. This is a relatively large force for practical and biological purposes in that it is comparable to the force generated by the most powerful molecular motors \cite{smith2001bacteriophage}, and about corresponds to the onset of the DNA overstretching transition observed in force spectroscopy~\cite{smith1996overstretching}. Second, at forces of $100$pN and beyond, the presence of knots is associated with significant slowing downs of the unzipping process depending on the interplay of knot topology and driving force.

\subsection{Effect of the unzipped strand interfering with the knot}

A noteworthy aspect of Fig.~\ref{Fig4} is the noticeable heterogeneity of the unzipping traces at $f=100$pN and $150$pN. For instance, over the five $4_1$ traces collected at $f=100$pN, the time required to reach the $n(t)=400$ mark can range from $1.2\cdot 10^6\tau_{MD}$ to $3.4\cdot 10^6\tau_{MD}$, a threefold ratio. For comparison, at $f=50$pN, the same ratio is only 1.02.

Visual inspection of the unzipping trajectories revealed that the heterogeneity is not only due to the presence of the knot but also to the hindrance arising from the unzipped ssDNA strand on the {\it cis} side becoming entangled with the knotted region. The effect is illustrated in Fig.~\ref{Fig5}, which presents typical DNA conformations on the {\it cis} side of the pore.

As illustrated, the knotted region typically leans against the pore entrance at the smallest considered force, $f=50$pN. However, at $f=100$pN and 150pN, the knot is often not in direct contact with the pore but is kept at a finite distance from it by the {\it cis} unknotted strand that wraps around the dsDNA stem immediately below the knot.
These wrappings originate from unwinding the unzipped DNA, which imparts a torque to the translocating molecule that accumulates at sufficiently high force when torsional stress is introduced faster than the relaxation dynamics can dissipate it.

Like those of  Fig.~\ref{Fig5}, the wrapped conformations inevitably offer a multi-tier hindrance to nanopore unzipping. The translocating dsDNA experiences the combined friction from the knot and the wrapped unzipped filament to a degree that depends on the tightness and number of turns of the latter, thus increasing the heterogeneity of the unzipping process.

\subsection{Knot dynamics}

We next considered the sliding dynamics of the knots along the {\it cis} portion of the DNA chain, which we addressed by tracking in time the nucleotide indices corresponding to the two ends of each knot. We employed the method of ref.~\cite{tubiana2011probing}, which uses a bottom-up search scheme to identify the shortest segment of a chain that, once closed with a suitable arc, yields a ring with the sought knot topology~\cite{tubiana2018kymoknot}.

\begin{figure*}[h]
\centering 
\includegraphics[width=2\columnwidth]{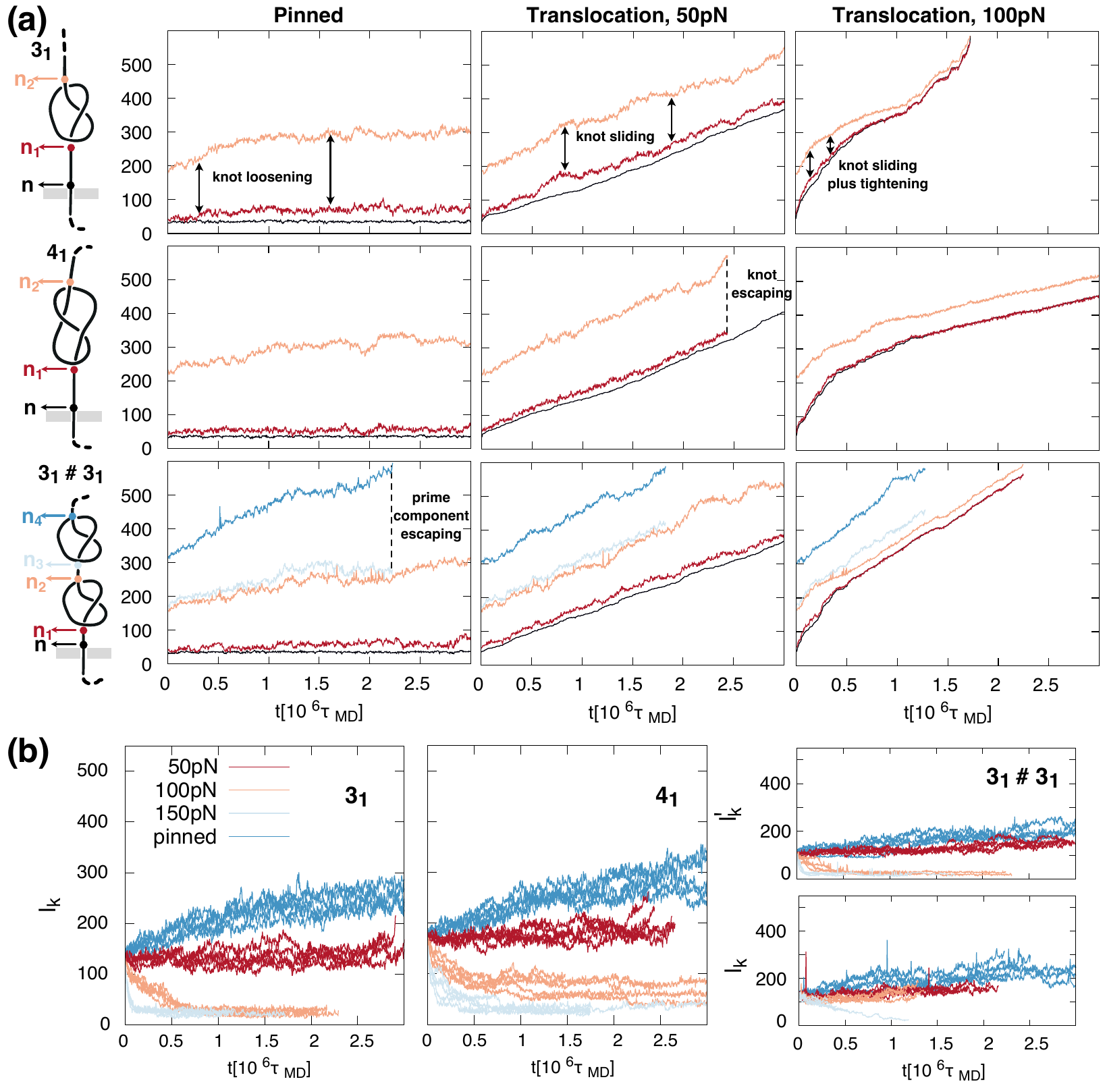}
\caption{(a) From top to bottom, three rows show the typical evolution of the contour positions of $3_1$, $4_1$, and $3_1\#3_1$ knots in different setups.
Sketches on the left provide the legend for the plotted nucleotide indices corresponding to the knot ends, $n_1, n_2$ for $3_1$ and  $4_1$ knots, and $n_1, n_2, n_3, n_4$ for the $3_1\#3_1$ knot. The $n(t)$ trace marks the index of the nucleotide at the pore entrance (or equivalently the number of translocated nucleotides, as in previous figures).
The first column is for a setup where a base inside the pore is kept pinned. The second and third columns represent translocation cases at 50pN and 100pN, respectively. The traces in panel (b) illustrate the time evolution of the knot length, $l_k$, for $3_1$, $4_1$ topologies, and for each of the two prime components for the $3_1\#3_1$ topology, $l_k$ and $l_k^\prime$. Each plot shows the pinned case, as well as 50, 100, 150pN pulling forces. The traces of five independent trajectories are shown for each case.}
\label{Fig6}
\end{figure*}

Fig.~\ref{Fig6}a illustrates the typical evolution of the contour positions of $3_1$, $4_1$ $3_1\#3_1$ knots for different forces. As indicated in the accompanying sketches, the $n_1$ and $n_2$ traces indicate the nucleotide indices of two ends of $3_1$ and $4_1$ knots and of the first (pore proximal) component of the $3_1\#3_1$ composite knot. The indices for the second component of the composite knot are instead indicated as $n_3$ and $n_4$. Additionally, the plots in Fig.~\ref{Fig6}a show the traces of the index of the nucleotide at the pore entrance, $n$.

The data in Fig.~\ref{Fig6}a allows for tracking various quantities of interest as a function of time, $t$. For instance, $n(t)$ is directly informative of the progress of the translocation/unzipping process. In contrast, the contour distance $n_1(t) - n(t)$ conveys how much the knotted region stays close to the pore during unzipping. In addition, the contour lengths of the prime knotted components are given by  $l_k=n_2(t)- n_1(t)$ and $l_k^\prime=n_4(t)- n_3(t)$ and are shown in Fig.~\ref{Fig6}b for the five independent trajectories of the considered cases.

\subsubsection{Knot evolution in pinned DNA chains.}

The first column in Fig.~\ref{Fig6}a is for the case where the ssDNA end inserted in the pore is not subject to a translocating force but is held in place by pinning a nucleotide inside the pore. The evolution of the pinned knotted configurations covers a time span of 3$\cdot 10^6\,\tau_{MD}$, comparable to the typical duration of unzipping processes at 100pN.
This case serves as a term of reference. Specifically, it establishes how the knotted DNA region evolves from its initial moderately tight state in the presence of the pore and slab but without any interference from a concurrent translocation/unzipping process and without mechanical tension propagating from the pore.
The traces of the pinned case show a systematic increase in knot lengths across all three considered topologies; see also Fig.~\ref{Fig6}b for $l_k$ and $l_k^\prime$.
The progressive loosening of knots reduces the system's bending energy compared to the initial state, where knotted components are moderately tight ($\sim$150bp) and significant curvature is thus packed into relatively short dsDNA stretches. The expansion of the knot is visibly asymmetric at the two ends because the knot cannot penetrate inside the pore and can only expand on the {\it cis} side.

The evolution of the $3_1\#3_1$ case is particularly interesting. The expansion is slowest for the first component (the one proximal to the pore), which is doubly constrained, being flanked by the slab and the pore on one side and the second $3_1$ knot on the other. The second knotted component, pushed by the first one, eventually reaches the free DNA end and thus becomes untied. From this point, the dynamics proceeds with the remaining $3_1$ knot, which reaches about the same size at the end of the simulated trajectory as the isolated $3_1$ knot, about 200bp, see also Fig.~\ref{Fig6}b.

We conclude that knots in pinned DNA chains can evolve substantially, expanding and becoming untied over timespans comparable to the entire unzipping process at $f=100 $ pN.

\subsubsection{Knot evolution during unzipping.}
The above dynamics is qualitatively modified when the pinning constraint is removed, and the DNA is forced to unzip by the driven translocation through the narrow pore.

The middle column of Fig.~\ref{Fig6}a is for $f=50$pN. In the $3_1$ case, the  $n(t)$, $n_1(t)$, and $n_2(t)$ traces are overall parallel, with $n_1$ staying close to $n$ at all times. These facts indicate that the $3_1$  knot remains close to the pore entrance throughout unzipping and maintains its initial moderately tightened state ($l_k\sim 150$bp) as (from the relative "perspective of the {\it cis} chain") it slides along the dsDNA contour at approximately constant velocity.

For the $4_1$ and $3_1\#3_1$ cases, the knots remain close to the pore entrance, their lengths slightly increase over time, albeit to a lesser extent than for the pinned case, with the $4_1$ reaching $l_k\sim 200$bp before escaping, and $3_1\#3_1$ reaching $l_k\sim 120$bp and $l_k^\prime\sim 150$bp for its prime components, Fig.~\ref{Fig6}b.

Increasing the force to $f=100$pN introduces radical changes to knot evolution and sliding dynamics, as seen in the rightmost plots of Fig.~\ref{Fig6}a. The $3_1$ knot exhibits a substantial tightening at the pore entrance, and so does the first $3_1$ component of the composite knot. Both values reach a stationary value of $l_k\sim 25$bp, Fig.~\ref{Fig6}b. Instead, the length of the second component of the composite knot appears to be only modestly affected, with $l_k^\prime$ fluctuating over values of $\sim 120$bp. Interestingly, the length of the $4_1$ knot also decreases with time, going from 200 nucleotides at $t=0$ to 70 at $t=3\cdot 10^6\tau_{MD}$, Fig.~\ref{Fig6}b, but never reaching the tightness observed at the late translocation stages of the $3_1$ knot.

Finally, at 150pN, the lengths of the $3_1$ knot and the first $3_1$ component of the composite knot both reach a similar asymptotic $l_k\sim 25$bp value as the ones of 100pN, but at a much faster pace, Fig.~\ref{Fig6}b. At this force, the $4_1$ knot can become tighter than 100pN, reaching an asymptotic value of $l_k\sim 30$bp, Fig.~\ref{Fig6}b.

The results clarify that the two dynamical regimes discussed for Fig.~\ref{Fig3}b are directly connected to the degree of tightness of the knot.
In fact, the $n(t)$ traces for $f=100$pN of Fig.~\ref{Fig6}a indicate that unzipping of the chains does not proceed at a constant pace but progressively slows down. The latter occurs in correspondence with the knot length reduction, conveyed by the close approach of the $n_1(t)$ and $n_2(t)$ curves.

The slow down, as well as its dependence on the applied force and knot type, is analogous to the topological friction found in translocating knotted chains without unzipping, shown in Refs.~\cite{rosa2012topological,suma2015pore} and shown for a dsDNA in Fig.~\ref{Fig1}. Similarly to these cases, the knot slows down the process but does not necessarily halt it entirely, as the chain can still slide on its knotted contour unless the dynamics is jammed by extreme knot tightening. The degree of tightening and, in turn, the associated hindrance depends on the applied force and the knot characteristics, which can change how the tension force propagates along the chain on the {\it cis} side.

\section*{Conclusions}

We used molecular dynamics simulations to study the nanopore unzipping of knotted DNA.
In our study, we considered dsDNA filaments of about 500bp prepared with different types of prearranged moderately tightened knots, namely the unknot (the trivial knot), $3_1$, $4_1$, and $3_1\#3_1$ knots. The filaments were unzipped by pulling one single-stranded terminus through a narrow pore at three different forces, $f=50$pN, 100pN, and 150pN. The progress of the unzipping process was characterized by analyzing the temporal traces of the number of translocated (hence unzipped) nucleotides and by tracking the position and length of the knotted region along the DNA contour.

The comparative analysis of the unzipping process across the considered knot types and forces enabled us to establish three main results. First, the DNA unzipping process at sufficiently low forces is virtually unaffected by the presence of knots. In fact, at $f=50$pN, the translocation traces of all three knot types were practically superposable to those of unknotted DNAs.
Second, increasing the force to $f=100$pN and 150pN caused knotted DNAs to unzip significantly more slowly and heterogeneously than unknotted ones. The highest hindrance was observed for $4_1$-knotted filaments, whose average unzipping at $f=150$pN was four times slower than the unknot. The corresponding dispersion of unzipping times was also substantial, accounting for a three-fold time difference between the slowest and fastest trajectories out of a set of five. Finally, analyzing the knotted DNA structure close to the pore revealed that the observed hindrance to unzipping involves at least two concurrent mechanisms: (i) the topological friction arising from the DNA chain sliding along its tightly knotted contour and (ii) the friction caused by the newly-unzipped {\it cis} DNA strand wrapping around the double-stranded DNA region between the knot and the pore.

The above results have implications in various physical and biological contexts. Because knots are statistically inevitable in sufficiently long DNA filaments, clarifying the impact of such forms of entanglement on how DNA unzips is relevant for polymer physics, particularly for developing predictive models for the complex force-dependent response of such processes. From the applicative point of view, the system and results discussed here could be used in prospective nanopore-based single-molecule unzipping experiments on long (hence knot-prone) DNAs, from interpreting the ionic current traces to designing such setups. Finally, DNA nanopore unzipping can be regarded as a gateway to elucidating the physical processes occurring {\it in vivo}, where genomic DNA is unzipped and translocated by various enzymes. It would thus be interesting to extend future considerations to DNA lengths and force regimes that match those relevant for  {\it in vivo} DNA transactions as closely as possible.

\section*{Model and numerical methods}

We used a coarse-grained model of DNA,  oxDNA2~\cite{oxdna1,oxdna2,oxdna3}, to simulate double-stranded DNA filaments of about 500 base pairs (bp). Each nucleotide is treated as a rigid body with three interaction centers. The potential energy describing the interactions between nucleotides accounts for the chain connectivity, stacking effects, excluded volume interactions, twist-bend coupling, base pairing (with sequence-averaged binding interactions), and screened electrostatic interactions.
The system was evolved with Langevin dynamics simulations using the LAMMPS  simulation package~\cite{LAMMPS,henrich2018}. The temperature was set to $T=300$K, and the monovalent salt concentration defining the Debye-Hueckel potential was set to 1 M NaCl.
Other model parameters were set to the default values of the LAMMPS oxDNA2 implementation, except for the friction coefficient set $\gamma=5$ as done in Ref.~\cite{suma2023nonequilibrium}. We use a timestep of $0.01$ $\tau_{MD}$. The longest duration of the simulations was $3.5 \cdot 10^6\tau_{MD}$.

Translocation was driven by a longitudinal force, $f=50,100,150$pN, acting exclusively on the DNA segment inside the pore and equally distributed among the nucleotides in the pore. This technical expedient is adopted to keep the driving force constant. The DNA strands have excluded volume interactions with a slab with an embedded cylindrical pore; see SI of Ref.~\cite{suma2023nonequilibrium} for the potential. The pore depth is 8.52nm, while the diameter is 1.87nm for unzipping simulations and 4.25nm for translocation simulation without unzipping shown in Fig.~\ref{Fig1} using the same initial setup described hereafter for unzipping translocation. Note that 1.87nm is a diameter sufficient to allow only a single ssDNA strand to pass at a time inside the pore, while 4.25nm is sufficient to allow a dsDNA strand to pass, but not a knot, composed by $\ge 3$ strands, which is forced to remain in the {\it cis} side of the pore.

To produce the initial conformation, we used an analogous scheme to Ref.~\cite{suma2015pore}: we first employed a  Monte Carlo scheme to sample equilibrated configurations of coarse-grained semi-flexible chains with thickness, contour length, and persistence length corresponding to double-stranded DNA filaments of 500 bp. At the front of the chain, a tightened knot was attached of three different types, $3_1$, $4_1$, $3_1\#3_1$, taken from simulations of Ref.~\cite{suma2015pore}, and long about 50bp. For the $0_1$ unknotted case, we didn't add anything.

The knotted terminus was then attached to a 40-base lead already threaded through the pore.
The configuration was subsequently relaxed using an intermediate fine-grained model, see Ref.~\cite{suma2017pore} for the specifics, by pinning one nucleotide inside the pore. During this relaxation the initially tight knotted components expand to about 150bp to lower the bending energy. The conformation was  then mapped to the oxDNA2 representation of double-helical DNA with the tacoxDNA package~\cite{suma2019tacoxdna}, with the lead inside the pore mapped into a  single-stranded DNA. The whole chain was again briefly relaxed by pinning one nucleotide inside the pore and letting the system evolve  for a timespan of $200\tau_{MD}$.

The relaxed filaments were translocated and unzipped by pulling the ssDNA stretch inside the pore with a constant total force, $f$.

A resulting initial conformation is shown in Fig.~\ref{Fig2}a.
At variance with Ref.~\cite{suma2023nonequilibrium}, here we show the translocation process for this configuration instead of unzipping the first 200bp bases, as our main interest is to study the knot positioning and effects.
Five different Monte Carlo-generated configurations were used for each topology, and their sequence composition was also randomly picked at the oxDNA fine-graining step.
The resulting conformations for the unknot and the three knot types are displayed in Fig.~\ref{Fig2}b during translocation.

Detection of knots was carried out using the software KymoKnot~\cite{tubiana2018kymoknot}. From a mathematical point of view, knots are rigorously defined only for circular chains. Accordingly, to establish the knotted state of an open chain, it is necessary to close it into a ring by bridging its terminals with a suitable auxiliary arc~\cite{tubiana2011probing}. This step was carried out with the so-called minimally-interfering closing procedure, which selects the auxiliary arc that adds the least possible entanglement to the open chain. After closure, the knotted state of the chain is established using the standard Alexander determinants, which suffice to pinpoint the chain's topological state unambiguously at our considered contour lengths and degree of confinement. This way, we assign a definite topological state to each configuration sampled in the MD trajectory.

\vskip 0.5cm

\section*{Acknowledgements}

This study was funded in part by the European Union - NextGenerationEU, in the framework of the PRIN Project "The Physics of Chromosome Folding" (code: 2022R8YXMR, CUP: G53D23000820006) and by PNRR Mission 4, Component 2, Investment 1.4\_CN\_00000013\_CN-HPC: National Centre for HPC, Big Data and Quantum Computing - spoke 7 (CUP: G93C22000600001). The views and opinions expressed are solely those of the authors and do not necessarily reflect those of the European Union, nor can the European Union be held responsible for them.

\bibliography{bibliography.bib}

\end{document}